# Classical theory of the hydrogen atom

**Sergey A. Rashkovskiy**

*Institute for Problems in Mechanics of the Russian Academy of Sciences, Vernadskogo Ave., 101/1, Moscow, 119526, Russia*

*Tomsk State University, 36 Lenina Avenue, Tomsk, 634050, Russia*

*E-mail: rash@ipmnet.ru, Tel. +7 906 0318854*

**Abstract** It is shown that all of the basic properties of the hydrogen atom can be consistently described in terms of classical electrodynamics instead of taking the electron to be a particle; we consider an electrically charged classical wave field, an "electron wave", which is held in a limited region of space by the electrostatic field of the proton. It is shown that quantum mechanics must be considered to be not a theory of particles but a classical field theory in the spirit of classical electrodynamics. In this case, we are not faced with difficulties in interpreting the results of the theory. In the framework of classical electrodynamics, all of the well-known regularities of the spontaneous emission of the hydrogen atom are obtained, which is usually derived in the framework of quantum electrodynamics. It is shown that there are no discrete states and discrete energy levels of the atom: the energy of the atom and its states change continuously. An explanation of the conventional corpuscular-statistical interpretation of atomic phenomena is given. It is shown that this explanation is only a misinterpretation of continuous deterministic processes. In the framework of classical electrodynamics, the nonlinear Schrödinger equation is obtained, which accounts for the inverse action of self-electromagnetic radiation of the electron wave and completely describes the spontaneous emissions of an atom.

## 1 Introduction

Modern quantum mechanics arose as a result of attempts to explain the discrete optical spectrum of the hydrogen atom.

The mechanistic model of the hydrogen atom suggested by N. Bohr (1913) became the first successful attempt, which was based on Planck's idea of quantization of the energy of the harmonic oscillator (which is considered to be the simplest model of an atom) and Einstein's idea of the quantization of radiation itself. The first of these ideas was the cornerstone of an explanation of the spectrum of equilibrium thermal radiation, while the second explained the laws of the photoelectric effect, which at that time could not be explained by classical electrodynamics.

Bohr's mechanistic model of the hydrogen atom was based on the following representations: (i) Rutherford's planetary model of the atom, (ii) the electron as a point charged particle as described by classical mechanics; (iii) discrete allotted stationary states of the atom, which correspond to the allotted (allowed) electron orbits; the remaining electron orbits predicted by classical mechanics do not exist; (iv) being in the stationary state, the atom does not radiate, and thus, it can reside in this state indefinitely; (v) an atom can jump from a stationary state with a lower energy into a stationary state with a higher energy only as a result of the absorption of a quantum of energy $\hbar\omega$, which is equal to the energy difference in the levels between which the transition occurs (induced transition); and (vi) the atom can spontaneously jump from a stationary state that has higher energy to a stationary state that has lower energy; in this case, it will emit a quantum of energy $\hbar\omega$, which is equal to the energy difference in the levels between which the transition occurs.

As a result of the formalization of this theory (A. Sommerfeld), the old quantum mechanics emerged, which allowed successful solution of a large number of problems, including discovering the properties of the electron, such as the spin. However, the assumptions and results of the old quantum theory contradicted classical mechanics, on which they were based.

Attempts to resolve these contradictions have ultimately led to the creation of wave mechanics (although the term "wave mechanics" is now almost not used; it appears that it better reflects the essence of this theory compared with the traditional term "quantum mechanics"), in which the basis, de Broglie's hypothesis (1924), argues that each electron has a wave-like behaviour and its corresponding wave length is related to the electron momentum by the ratio introduced earlier by Einstein for light quanta - photons.

The discovery of the Schrödinger equation (1926) was the first success in this direction. This equation described the electron in a hydrogen atom with the help of a continuous wave field $\psi$. This equation has a discrete set of stationary solutions and a corresponding discrete set of eigenvalues, which coincides with the discrete energy levels that were predicted by the old quantum theory. This arrangement means that the problem of describing the hydrogen atom is reduced to the solution of the Schrödinger equation and finding its eigenvalues.

However, the connection of the continuous wave field $\psi$, as described by the Schrödinger equation, with the electron as a particle, has remained unclear. The first interpretation of the wave field $\psi$ was suggested by E. Schrödinger, which considered it to be a real physical field. However, Schrödinger's point of view has been heavily criticized first by N. Bohr, W. Heisenberg and P. Dirac, and also by their followers, and the so-called Copenhagen interpretation, based on the wave-particle duality, emerged and became the official point of view. The unquestioned authority of N. Bohr contributed to this perspective to a large extent, while his mechanistic representations about the existence of discrete stationary states of the hydrogen atom and jump-like transitions between those states, which are accompanied by the emission or absorption of a quantum of energy $\hbar\omega$, have become considered to be a reference to which any theory that describes the atom must conform. The confidence that Schrödinger's interpretation will inevitably contradict Planck's theory of thermal radiation, which demands the quantization of energy, was another argument against Schrödinger's interpretation.

From a formal point of view, the Copenhagen approach reduces to the viewpoint that we first solve the Schrödinger equation (or other wave equation) and find a continuous wave field $\psi$, which afterward is interpreted from the standpoint of the original Bohr's mechanistic theory, in other words, from the viewpoint of the existence of discrete states, jump-like transitions and the probability of the location of an electron considered to be a particle in these states.

In other words, in spite of a fundamentally new way of describing the electron, which wave mechanics suggests, the interpretation of the results remained old, i.e., mechanistic.

Almost immediately, wave mechanics ran into difficulty, which was overcome only in the framework of quantum electrodynamics, i.e. in the result of "second quantization". The spontaneous emission of an atom was the first such challenge.

It is accepted that spontaneous transitions are not explainable within the framework of the Schrödinger equation, in which the electron energy levels are quantized but the electromagnetic field is not. Thus, in the absence of a quantized electromagnetic field, the excited state of the atom cannot decay to the ground state. To explain spontaneous transitions, quantum mechanics must be extended to quantum electrodynamics, wherein the electromagnetic field is quantized at every point in space. In quantum electrodynamics, the electromagnetic field has a ground state, the QED vacuum, which can "mix with the excited stationary states of the atom". As a result of this interaction, the "stationary state" of the atom is no longer a true eigenstate of the combined system of the atom plus the electromagnetic field. Thus, it is accepted that the *spontaneous emission in free space depends on vacuum fluctuations to get started* [1].

There are attempts to explain the quantum phenomena within the framework of so-called semi-classical theory, in which only the states of the atom are quantized, while light is considered to be a classical Maxwell field [2-12].

Although this approach leads to correct results in some cases, it cannot be considered to be consistent because it denies the wave-particle duality of light but retains it in relation to the electrons.

The attempt to resolve this contradiction is made by the author in a series of papers, one of which is the present paper.

In previous papers of this series [13-15], it was shown that there is no need to introduce physical objects such as photons for the explanation of many quantum phenomena: the discrete character of the interaction of light with a detector (matter) observed in the experiments can be explained in the framework of classical electrodynamics, when light is considered to be a continuous classical electromagnetic field.

Considering the formal similarity of the wave properties of light and electrons as well as the conclusions of previous studies [13-15], it was shown [16] that for a consistent explanation of many "quantum" effects, we should abandon the idea of electrons as particles and instead consider the classical physical wave field – an electron wave having unusual (compared to an electromagnetic field) properties: continuously distributed in space are the electric charge, the internal angular momentum and, associated with it, the internal magnetic moment.

This arrangement means that each elementary volume of space $dV$, which is filled with the electron field, possesses an electric charge, angular momentum and magnetic moment [16]. This angular momentum and magnetic moment distributed in space are not connected with the movement of any charged particles and are a fundamental property of the electron field itself, similar to matter that is continuously distributed in space. In particular, this property is manifested in that the gyromagnetic ratio for the electron field (the ratio of the magnetic moment of any volume of the space filled with the electron field to the angular momentum of the same volume) is 2 times greater than the gyromagnetic ratio that is connected with the motion of the electric charges (either a point or continuously distributed in space).

From this point of view, the Dirac equation that describes the electron wave field plays the same role as Maxwell equations for the electromagnetic field, i.e., it is a field equation. Other well-known wave equations (Klein-Gordon, Pauli and Schrödinger) are simplified forms of the Dirac equation that describes the electron field in various approximations. As shown in [16], the electron field described by the Dirac equation should be considered to be a classical field that is similar to the classical electromagnetic field, i.e., it is continuous in space and time.

Such a point of view on the "electron" avoids many of the paradoxes that are peculiar to the Copenhagen interpretation and allows explaining in a natural way many of the features of its behaviour, in particular, the "wave-particle duality" in its various manifestations (e.g., in the double-slit experiments) and the Born rule, which is a mathematical expression of the wave-particle duality. It is also consistent with the classical explanation of "the Heisenberg uncertainty principle".

Adhering to the point of view in [13-16], we are arriving at considering a new viewpoint on the structure of the atom and giving a classical explanation for the well-known atomic phenomena.

In this article, I will show that all of the main properties of the hydrogen atom, such as the following

- the existence of stationary states, in which the atom does not radiate electromagnetic waves;
- the stability of the atom;
- the nature of the spontaneous emission;
- the discrete spectrum of spontaneous emissions;
- the lack of some lines in a spectrum of spontaneous emission (which is usually interpreted as "the forbidden transitions");
- the width of the spectral lines of spontaneous emission;

and many others have a simple and evident explanation within the framework of classical field theory (classical electrodynamics) and do not require any quantization.

I will show that there are no discrete states and discrete energy levels of the atom: the energy of the atom and its states change continuously and, therefore, there are no jump-like transitions of the atom between "discrete levels". I will show that the spontaneous emission of an atom is a continuous classical process that is completely described by classical electrodynamics. In this paper, it will be shown that the wave equations that describe the electron wave (Schrödinger, Klein-Gordon, Pauli, Dirac) and electromagnetic field (Maxwell) as classical fields are sufficient for the consistent description of the main optical properties of the hydrogen atom.

## 2  Electromagnetic analogy in quantum mechanics

2.1  Infinite potential well

The infinite potential well is a simple one-dimensional analogue of the hydrogen atom; it is capable of holding a "quantum particle" in a limited region of space. For any "finite motion", as in this case, the Schrödinger equation has a discrete set of stationary solutions that are described by the eigenfunctions that correspond to the definite eigenvalues. From a mathematical point of view, these solutions correspond to standing waves. According to the generally accepted interpretation of the solutions of the Schrödinger equation, the eigenvalues are considered to be the energy of a quantum particle (for example, "electron"), and next, it is concluded that the "electron", being placed in such a potential well, can only have the discrete set of energies [17] $E_n = \frac{\pi^2 \hbar^2}{2 m_e a^2} n^2$, where $n = 1,2,3, ...$, $m_e$ is the mass of an "electron", and $a$ is the width of the potential well. A rectangular "potential box" with dimensions $a_1 \times a_2 \times a_3$ is a three-dimensional generalization of the infinite potential well. The Schrödinger equation for the "electron" in such a "potential box" also has a discrete set of eigenvalues $E_{n_1,n_2,n_3} = \frac{\pi^2 \hbar^2}{2m_e} \left( \frac{n_1^2}{a_1^2} + \frac{n_2^2}{a_2^2} + \frac{n_3^2}{a_3^2} \right)$, where $n_1, n_2, n_3 = 1,2,3, ...$, based on which it is concluded that the "electron", being placed in a "potential box", can have only the discrete set of energies $E_{n_1,n_2,n_3}$. At the same time, the corresponding wave functions (eigenfunctions of the Schrödinger equation) $\psi_{n_1,n_2,n_3} = \sqrt{\frac{8}{a_1 a_2 a_3}} \sin \frac{\pi n_1 x}{a_1} \sin \frac{\pi n_2 y}{a_2} \sin \frac{\pi n_3 z}{a_3}$ are used to calculate the probability of finding the particle at different points in space; thus, the value of $\left| \psi_{n_1,n_2,n_3} \right|^2 dxdydz$ is interpreted as the probability of finding the "electron" in a state $(n_1, n_2, n_3)$ in a small region $dxdydz$ in the neighbourhood of the point $(x, y, z)$.

In electrodynamics, a cavity resonator, which represents a rectangular cavity with perfectly reflecting walls, is a direct analogue of the "quantum" rectangular "potential box". Maxwell's equations for classical electromagnetic waves enclosed in a cavity resonator have specific ("stationary") solutions in the form of standing waves, which correspond to the frequencies (eigenvalues of Maxwell equations) $\omega_{n_1,n_2,n_3}^2 = \pi^2 c^2 \left( \frac{n_1^2}{a_1^2} + \frac{n_2^2}{a_2^2} + \frac{n_3^2}{a_3^2} \right)$, where $c$ is the speed of light. [18]

Such cavity resonators are widely used in various technical applications [19], in particular, in microwave devices, and the electromagnetic waves held by them are always considered to be classical Maxwell fields. No one would ever consider them to be "microwave particles", the energy of which is quantized, in other words, can take on only certain discrete values $E_{n_1,n_2,n_3} = \hbar \omega_{n_1,n_2,n_3}$. Moreover, even if in such a cavity resonator, there is only one eigenmode, which corresponding to a certain eigenfrequency $\omega_{n_1,n_2,n_3}$ that is excited; the energy of the electromagnetic wave can take any continuous value, which will be determined by the wave amplitude. In general, any electromagnetic wave field can be created in such a cavity resonator, and this field can be represented as a superposition of the eigenmodes of the cavity resonator that correspond to the different eigenfrequencies $\omega_{n_1,n_2,n_3}$.

The question arises as to why we consider the electromagnetic wave in a rectangular cavity resonator to be a classical wave field while the electron wave in a similar cavity resonator - a rectangular "potential box" - is considered to be a "quantum particle", e.g., an "electron"? In both cases, the object in the cavity is described by continuous functions of spatial coordinates and time, which are the solutions of partial differential equations – the field equations.

If we consider, instead of an "electron" ("quantum particle"), an electron wave, which is regarded as a classical wave field that is described approximately by the Schrödinger equation [16], we can then take a fresh look at the solution of the Schrödinger equation for a rectangular "potential box" that allows us to avoid the contradictions and paradoxes that are inherent in the Copenhagen interpretation of these solutions. In fact, assuming that an electron wave is a classical wave field [16] and considering the solutions of the Schrödinger equation, one concludes that in the cavity resonator (i.e., in a "quantum" rectangular "potential box"), there are the standing electron waves which correspond to the eigenmodes of this cavity resonator. These eigenmodes have certain frequencies $\omega_{n_1,n_2,n_3} = E_{n_1,n_2,n_3}/\hbar$, which are the eigenfrequencies of the cavity resonator (the eigenvalues of the Schrödinger equation), and they are described by the eigenfunctions of the Schrödinger equation. In general, any of the electron wave fields (not only the standing waves) can be created in the cavity resonator, and the arbitrary electron wave field in the cavity resonator can be represented as a superposition of the eigenmodes of the resonator:

$$\psi(x,y,z,t) = \sum_{n_1,n_2,n_3} c_{n_1,n_2,n_3} \psi_{n_1,n_2,n_3}(x,y,z) \exp(-i\omega_{n_1,n_2,n_3} t) \quad (1)$$

where $c_{n_1,n_2,n_3}$ are the amplitudes of the corresponding eigenmodes, and the eigenfunctions $\psi_{n_1,n_2,n_3}(x,y,z,t)$ are orthonormal:

$$\int_0^{a_3}\int_0^{a_2}\int_0^{a_1} \psi^*_{n'_1,n'_2,n'_3}(x,y,z)\psi_{n_1,n_2,n_3}(x,y,z)dxdydz = \delta_{n_1 n'_1}\delta_{n_2 n'_2}\delta_{n_3 n'_3} \quad (2)$$

As follows from the self-consistent system of equations of the electron-electromagnetic field [16], the value $-e|\psi|^2$ in the Schrödinger approximation is equal to the electric charge density of the electron field, where $e = 1.6021766208\,(98)\times10^{-19}$ C is the elementary charge. If the total charge of the electron field in the cavity is equal to $-e$ (the "single-electron" field), then it follows from (1) and (2) that

$$\sum_{n_1,n_2,n_3} |c_{n_1,n_2,n_3}|^2 = 1 \quad (3)$$

In quantum mechanics, this condition is obtained based on a probabilistic interpretation of the wave function, which allows interpreting the value $|c_{n_1,n_2,n_3}|^2$ as the probability of finding the "electron" in a quantum state that is characterized by the quantum numbers $(n_1, n_2, n_3)$. Considering the electronic wave as a classical charged wave field [16], we arrive at another interpretation of the parameters $|c_{n_1,n_2,n_3}|^2$. Indeed, the value $-e\sum_{n_1,n_2,n_3}|c_{n_1,n_2,n_3}|^2$, by definition, equals the electric charge of the entire electron wave that is enclosed in the cavity resonator. Then, the value $-e|c_{n_1,n_2,n_3}|^2$ is equal to the electric charge of the part of the electron wave that is in eigenmode $(n_1, n_2, n_3)$. In this case, condition (3) expresses the law of conservation of charge: the electric charge of an electron wave can be redistributed between the eigenmodes of the cavity resonator in such a way that its total charge remains equal to $-e$.

## 2.2 The Maxwell-Dirac isomorphism

The propagation of electromagnetic waves in a homogeneous but isotropic and stationary medium is described by Maxwell's equations, which for a monochromatic wave take the form [18]

$$\text{rot}\mathbf{E} = i\mu\frac{\omega}{c}\mathbf{H}, \ \text{rot}\mathbf{H} = -i\varepsilon\frac{\omega}{c}\mathbf{E} \tag{4}$$

where $\varepsilon$ and $\mu$ are the permittivity and permeability of the medium depending on the coordinates, and $\omega$ is the frequency of the electromagnetic wave.

As shown in [20], with

$$\varepsilon = 1 + \frac{e\varphi + m_e c^2}{\hbar\omega}, \ \mu = 1 + \frac{e\varphi - m_e c^2}{\hbar\omega} \tag{5}$$

where $\varphi$ is a function of the coordinates, and with the additional conditions

$$(\mathbf{E}\nabla)\varepsilon = 0, \ (\mathbf{H}\nabla)\mu = 0 \tag{6}$$

Maxwell's equations (4) can be written in the form of a stationary Dirac equation, which describes the "electron" in an electrostatic field $\varphi$.

This finding means that if we create a medium that has permittivity and permeability (5), then the classical (continuous) electromagnetic waves in it that satisfy the conditions in (6) will behave in the same way as an "electron" in the external electrostatic field $\varphi$. At the least, both of these electromagnetic waves and the "electron" will be described by the same functions and will have the same eigenfrequencies.

Note that it is impossible to create the medium in (5) using conventional materials; however, it can be created using metamaterials [21]. By creating this medium, we can simulate the "quantum effects" and even recreate the "quantum system" by using classical macroscopic systems and classical electromagnetic waves.

## 2.3 "Optical" atom

As an illustrative example, let us consider a hypothetical spherically symmetric medium that has permittivity and permeability (5) with $\varphi = e/r$, where $r$ is the distance from the centre of symmetry. Let us inject into this medium classical electromagnetic waves. As shown in [20,22], the waves that satisfy the conditions in (6) are described by the Dirac equation for the hydrogen atom. This arrangement means that in such a medium, the undamped standing electromagnetic waves can exist and correspond to only certain frequencies of the discrete spectrum, which are the eigenvalues of the Dirac equation. Such a medium will be an open volume resonator in which the classical electromagnetic wave is held due to the total internal reflection in the inhomogeneous dielectric medium [19]. From a mathematical point of view, such a medium is a complete analogue of the hydrogen atom because the standing electromagnetic waves in it will be described by the same functions and will have the same eigenfrequencies as the hydrogen atom. For this reason, we will call this medium the "optical atom".

Note that here we are considering the classical continuous electromagnetic waves in an open volume resonator, which can have any energy depending on their amplitude; the only frequencies of the standing waves are discrete. This arrangement is similar to the arrangement that we discussed in section 2.1 in relation to the cavity resonator. Despite the fact that the standing waves in the open volume resonator in their wave properties are similar to the "electron" in the hydrogen atom, they have a simple and intuitive classical sense, and thus, we do not need to explain their behaviours and properties, in particular the discrete spectrum of eigenfrequencies, by the wave-particle duality. Hence, the question arises as to whether we need the wave-particle duality to explain the properties of the hydrogen atom. To exclude any attempts to attract to an explanation of the properties of the "optical atom" such objects as "photons", note that we can choose the parameters of the medium (5) in such a way that all of the eigenfrequencies of such an open volume resonator will be, for example, in the microwave range, which excludes any doubts as to the classical nature, both for the open volume resonator itself and for the electromagnetic waves that are locked in it.

Here, it might be tempting to identify the real hydrogen atom with the "optical atom", in which a classical electromagnetic wave is held, and thereby to attempt to reduce the electron waves ("electrons") to electromagnetic waves. This idea was expressed in [20,22]. Despite the attractiveness of such an idea, we apparently should abandon it, at least from the literal identification of the electron and electromagnetic waves because the electromagnetic waves, unlike electron waves, do not carry an electric charge and, moreover, do not have internal angular momentum and magnetic moment continuously distributed in space [16].

## 3  What is an atom?

Using the representation in [16], I affirm and will further substantiate that the atom represents a positively charged nucleus, which holds around it a negatively charged electron wave that is continuously distributed in space, which is a true existing classical field. Because the electron wave is continuous, its total electric charge can take on any continuous values; however, due to the electrical neutrality of the atom, the total electric charge of the electron wave in the atom is always equal in magnitude to the electric charge of the nucleus. Thus, the "quantization" of the electric charge of the "electron" in the atom is explained by the discreteness of the electric charge of the atomic nucleus, which contains an integer number of protons.

Using the results of the previous section, one can also say that *the atom represents an open volume resonator* in which electron wave - a classical wave field - is "locked". The electron wave because it is electrically charged, is held in such a resonator by the electrostatic field of the nucleus. However, given the electromagnetic analogy, the electrostatic field of the nucleus can be conventionally considered to be a dielectric and magnetic medium in which the electron wave propagates. The permittivity and permeability (5) of the medium depends on the potential of the electrostatic field of the nucleus. Drawing an analogy with the "optical atom", it can be considered that the electron wave in an atom is held due to the "total internal reflection" on the inhomogeneities of the "dielectric medium".

In this paper, we will only consider a hydrogen atom. More complex "multi-electron" atoms will be discussed in the subsequent papers of this series.

Note that the proton, in contrast with the electron wave, is considered here to be a point electric charge. There is some inconsistency in such an approach, but this issue will be considered in the subsequent papers of this series.

From this point of view, the Dirac equation and its various approximations (Klein-Gordon, Pauli, Schrödinger) are the field equations that describe the behaviour of the electron wave in an open volume resonator - the hydrogen atom.

In this paper, I will consider the electron wave in the Schrödinger approximation [16], without accounting for its properties, such as spin.

The electron wave in the hydrogen atom in this approximation is described by a scalar wave function

$$\Psi(t, \mathbf{r}) = \exp(-i\omega_e t)\psi(t, \mathbf{r}) \qquad (7)$$

where the function $\psi(t, \mathbf{r})$ satisfies the Schrödinger equation for the hydrogen atom, while the frequency $\omega_e$ is related to the "mass of an electron" $m_e$ by the ratio

$$m_e = \hbar\omega_e/c^2 \qquad (8)$$

In this approach, the frequency $\omega_e$ is considered to be the primary characteristic of the electron wave, while the "mass of the electron" is an auxiliary value that is used for convenience in some (primarily in classical) applications [16].

The electric charge of the electron wave is continuously distributed in space around the proton with the density

$$\rho = -e|\psi|^2 \tag{9}$$

The total charge of the electron wave in an electrically neutral hydrogen atom is equal to $-e$, and thus,

$$\int |\psi|^2 dV = 1 \tag{10}$$

where the integral is taken over the entire space.

In quantum mechanics, this condition is interpreted as a normalization condition for the probability density $|\psi|^2$; however, we see here that condition (10) has a simple classical meaning and expresses only the electrical neutrality of the hydrogen atom.

The Schrödinger equation for the hydrogen atom has an infinite discrete set of stationary solutions

$$\psi_n(t, \mathbf{r}) = u_n(\mathbf{r}) \exp(-i\omega_n t) \tag{11}$$

where $u_n(\mathbf{r})$ and $\omega_n$ are the eigenfunctions and eigenvalues of the stationary Schrödinger equation. They can be found in any textbook on quantum mechanics, for example, in [17]. In this paper, a specific form of these solutions will not interest us, and thus, instead of the three usual "quantum numbers", we use simply the "number" of the stationary solution $n$.

The eigenfunctions $u_n(\mathbf{r})$ are orthonormal:

$$\int u_n(\mathbf{r}) u_k^*(\mathbf{r}) dV = \delta_{nk} \tag{12}$$

Each stationary solution (11) describes a standing electron wave in the open volume resonator, which is a hydrogen atom. Each of these standing waves is one of the eigenmodes of this resonator. If only one of the eigenmodes $n$ of the hydrogen atom is excited, then the whole electric charge of the electron wave $-e$ is concentrated in this mode. This view exactly explains the normalization (12) of the eigenfunctions.

A state of the atom in which only one of its eigenmodes is excited will be called here the pure state. Thus, the stationary solution (11) describes a pure state $n$.

In general, any number of eigenmodes can be simultaneously excited in the hydrogen atom. Due to the linearity of the Schrödinger equation, its general solution can be written as a superposition of specific solutions (11):

$$\psi(t, \mathbf{r}) = \sum_n c_n u_n(\mathbf{r}) \exp(-i\omega_n t) \tag{13}$$

where $c_n$ are the constants that determine the amplitudes of excitation of various eigenmodes.

A state of the atom in which at least two of its eigenmodes are excited simultaneously will be called here the mixed state.

Note that the terms "pure state" and "mixed state" are already "occupied" in quantum mechanics, but they have different meanings that are associated with the measurement procedure [17], which plays a fundamental role in the Copenhagen interpretation. In my opinion, such usage of these terms is unfortunate because it does not reflect the true meanings for these concepts.

The use of the terms "pure state" and "mixed state" for the states of the hydrogen atom as an open volume resonator, in which one or simultaneously several eigenmodes are excited, respectively, appears to be more reasonable and logical. For this reason, in this paper and in subsequent papers of this series, the terms "pure state" and "mixed state" will be used only in this sense.

Because the total charge of the electron wave in the hydrogen atom in the mixed state is equal to $-e$, using (9), (12) and (13) yields

$$\sum_n |c_n|^2 = 1 \tag{14}$$

In the probabilistic interpretation of quantum mechanics, the parameters $|c_n|^2$ are interpreted as the probability of "finding an electron in a quantum state $n$", and condition (14) is interpreted as a normalization condition for the probabilities.

Considering the electron wave as a classical charged field, we see that expression (14) expresses only the condition of electrical neutrality of the atom, while the parameters $|c_n|^2$ describe the distribution of the electric charge of the electron wave between the excited eigenmodes of the atom: the value $-e|c_n|^2$ is equal to the electric charge of the electron wave, which is contained in an eigenmode $n$.

## 4  Spontaneous emission of an atom

### 4.1  Spectrum and intensity of the spontaneous emission

Considering the electron wave to be a real physical field that has a spatially distributed electric charge, one can use classical electrodynamics for the calculation of the electromagnetic fields that are created by it.

Using this approach, we calculate the electric dipole moment of the electron wave in the hydrogen atom:

$$\mathbf{d} = -e \int \mathbf{r} |\psi|^2 dV \tag{15}$$

For the hydrogen atom that is in the mixed excited state, by substituting (13) into (15), we obtain

$$\mathbf{d} = \sum_n \sum_k c_n c_k^* \mathbf{d}_{nk} \exp(-i\omega_{nk} t) \tag{16}$$

where

$$\omega_{nk} = \omega_n - \omega_k \tag{17}$$

$$\mathbf{d}_{nk} = \mathbf{d}_{kn}^* = -e \int \mathbf{r} u_n(\mathbf{r}) u_k^*(\mathbf{r}) dV \qquad (18)$$

are constant vectors.

As shown below, the amplitudes of the excitations of the eigenmodes $c_n$ in a mixed state are time-dependent, but their change is slow compared with the rapidly oscillating factors $\exp(-i\omega_{nk}t)$. Assuming that there is the condition

$$|\dot{c}_n| \ll |\omega_{nk}| \qquad (19)$$

which will be proven below, one obtains approximately

$$\ddot{\mathbf{d}} = -\sum_n \sum_k \omega_{nk}^2 c_n c_k^* \mathbf{d}_{nk} \exp(-i\omega_{nk}t) \qquad (20)$$

According to classical electrodynamics [23], such a system creates electric dipole radiation, the intensity of which is given by

$$I = \frac{2}{3c^3} \overline{\ddot{\mathbf{d}}^2} \qquad (21)$$

where the bar denotes averaging over time.

Substituting (20) into (21), after averaging, we obtain

$$I = \sum_{\omega_{nk}} I_{nk} \qquad (22)$$

where

$$I_{nk} = \frac{4\omega_{nk}^4}{3c^3} |c_n|^2 |c_k|^2 |\mathbf{d}_{nk}|^2 \qquad (23)$$

is the intensity of the electric dipole radiation at a frequency of $\omega_{nk}$.

This finding is exactly the spontaneous emission of an atom, and it has a completely classical sense. Equation (23) for the intensity of the spontaneous emission of an atom was obtained in the framework of classical electrodynamics, although traditionally it is derived in quantum electrodynamics when using representations of the probabilities of jump-like transitions, the emission of the photons, photon polarizations, changes in the occupation numbers, and other factors. [24].

From the above derivation, all of the basic properties of the spontaneous emission follow.

The atom, which is in a pure state (11), does not radiate because in this case, the electron wave has a constant electric dipole moment. This circumstance implies that the pure states are the stationary states of the atom: the atom can be in these states indefinitely because it does not lose energy due to emission. In contrast, for an atom that is in a mixed state when at least two eigenmodes (11) are excited, the electrically charged electron wave has a nonstationary electric dipole moment (16), and according to classical electrodynamics, it radiates electromagnetic waves at the frequencies $\omega_{nk}$ for all $n$ and $k$, which correspond to the excited eigenmodes (11).

Because the hydrogen atom as an open volume resonator has the discrete spectrum of the eigenfrequencies $\omega_n$, the spectrum of the spontaneous emission $\omega_{nk}$ (17) is also discrete. Thus, the discreteness of the spectrum of spontaneous emission of an atom is not related to "jump-like transitions", and instead, it has a simple and clear explanation in the framework of classical electrodynamics.

In contrast to the traditional viewpoint that has been adopted in quantum mechanics, according to which the radiation is emitted in the form of discrete portions - quanta, we see that *the spontaneous emission of an atom is a continuous process,* and it continues as long as at least two eigenmodes of the atom remain excited simultaneously. As shown below, a change in the state of the atom occurs due to spontaneous emission, which leads to the gradual attenuation of spontaneous emission, unless the mixed excited state of the atom is not supported by an external influence. However, in any case, *the energy of the atom in the process of spontaneous emission changes continuously* with the rate (22).

In quantum mechanics, it is considered that an atom being in a stationary state can suddenly go to a lower energy level, while emitting an energy quantum. For an explanation of the reasons for such a transition, representations were formed that regard fluctuations of the QED vacuum, which "swings the electron" (*Zitterbewegung*); the electron is originally located on a stationary energy level, and this phenomenon induces its transition. Such representations underlie quantum electrodynamics and have led to the emergence of representations that regard the QED vacuum. The above analysis shows that an external "jolt" to the atom is not required to begin spontaneous emission; emission occurs continuously and begins immediately after the atom is excited, for some reason, into a mixed excited state.

Thus, we see that the spontaneous emission of an atom is devoid of any mystical sense. The emission is associated with the oscillations of a continuous in the space electric charge of the electron wave, which is held by the electrostatic field of the nucleus of the atom and occurs in full compliance with classical electrodynamics. From this point of view, *the atom is a microscopic classical dipole antenna that is continuously emitting electromagnetic waves*.

4.2 Allowed and forbidden "transitions"

In quantum mechanics, the absence of some lines in the spectrum of spontaneous emission is traditionally interpreted as the existence of "forbidden transitions".

To find the "forbidden transitions", special selection rules were formulated in quantum mechanics, which are based on Bohr's correspondence principle, but a physical explanation of why these "transitions" are forbidden is absent.

In the proposed model of the hydrogen atom that is considered to be an open volume resonator, holding the classical electron wave by the electrostatic field of the proton (or due to "total internal reflection" of the electron wave), there are no discrete states and, consequently, jump-like transitions, and thus, it makes no sense to talk about the allowed or forbidden "transitions". The proposed completely classical model of the atom allows us to easily understand why some of the expected spectral lines in the spectrum of the spontaneous emission are not observed. There is no need to enter any special "quantum" rules or assumptions: the result follows in a natural way from classical electrodynamics, and it is completely based on a classical picture.

As was shown above, an atom that is in a mixed excited state continuously emits electromagnetic waves of the discrete spectrum $\omega_{nk}$. The intensity of the radiation at a frequency $\omega_{nk}$ is determined by the expression (23), which follows from classical electrodynamics and depends on the parameter $|\mathbf{d}_{nk}|^2$. As follows from (23), for those frequencies $\omega_{nk}$ for which $|\mathbf{d}_{nk}|^2 = 0$, the intensity of the spontaneous emission is equal to zero. This circumstance means that the spectral lines that correspond to such frequencies will not be observed. From the condition $|\mathbf{d}_{nk}|^2 = 0$, purely mathematically [17] and without any special "quantum" interpretation, one can easily derive the selection rules that now no longer have the mystical sense that is attributed to them in quantum mechanics.

## 4.3 Damping due to spontaneous emission

The emission of the electromagnetic waves by the charged electron wave in a mixed excited state of an atom results in a loss in the energy that an electromagnetic wave carries away. In this case, the energy of the electron field decreases with the emission rate (22), (23). This change leads to a redistribution of the electron wave between the eigenmodes (11). From a mathematical point of view, this arrangement means that the parameters $c_n$ in the general solution (13) are not constant and instead vary with time: $c_n = c_n(t)$.

The linear Schrödinger equation cannot describe this process because its general solution has the form (13) with constant coefficients $c_n$. In particular, according to the Schrödinger equation, the energy of the electron wave

$$E \equiv \langle \psi | \widehat{H} | \psi \rangle = \sum_n \hbar \omega_n |c_n|^2 \qquad (24)$$

remains constant, where $\widehat{H} = i\hbar \frac{\partial}{\partial t}$.

This finding points to the limitations of the linear Schrödinger equation, which allows for the calculation of the source of the electromagnetic radiation (13) but does not allow for the calculation of the change in the state of the electron wave during the process of emission. For such a calculation, it is necessary to use a nonlinear equation of the electron field, which accounts for a reverse action of the emitted electromagnetic wave on the electron wave. The absence of such an equation in quantum theory is replaced by the second quantization, which implicitly contains such a nonlinear equation.

In my opinion, the second quantization is only a mathematical technique for the approximate solution of a nonlinear wave equation that describes the electron wave. As shown below, it is not necessary to use the method of second quantization for the description of spontaneous emission because all of the results can be obtained directly by solving a nonlinear field equation for the electron wave.

Let us first consider an approximate method of calculating the change in the state of the electron wave in the hydrogen atom during the spontaneous emission of electromagnetic waves. To obtain this goal, we use the law of conservation of energy

$$\frac{dE}{dt} = -I \qquad (25)$$

which, when accounting for expressions (22) and (24), takes the form

$$\sum_n \hbar\omega_n \frac{d|c_n|^2}{dt} = -\sum \omega_{nk} I_{nk} \qquad (26)$$

This equation should be solved while accounting for the law of conservation of charge of the electron field (14): in spontaneous emission, a redistribution of electric charge between the eigenmodes of the electron wave occurs, in such a way that the total electric charge in the hydrogen atom remains constant and equal to $-e$.

Let us consider the case in which the only two eigenmodes $n$ and $k$ of the electron wave in the hydrogen atom are excited. Such an electron wave is in a mixed state and will emit electromagnetic waves at a frequency of $\omega_{nk}$. In this case, the energy loss will occur at the rate $I_{nk}$, which will be accompanied by a redistribution of the electric charge of the electron wave between eigenmodes $n$ and $k$. For definiteness, we assume that $\omega_n > \omega_k$.

The law of conservation of energy (26) in this case has the form

$$\hbar\omega_n \frac{d|c_n|^2}{dt} + \hbar\omega_k \frac{d|c_k|^2}{dt} = -I_{nk} \qquad (27)$$

while the law of conservation of charge (14) can be written as

$$\frac{d|c_n|^2}{dt} + \frac{d|c_k|^2}{dt} = 0 \qquad (28)$$

Then, we obtain

$$\frac{d|c_n|^2}{dt} = -\frac{d|c_k|^2}{dt} = -\frac{I_{nk}}{\hbar\omega_{nk}} \qquad (29)$$

This condition expresses the balance of the electric charge of the electron wave in the hydrogen atom: in the process of spontaneous emission, the electric charge of the electron wave "crossflows" from the eigenmode that has a greater eigenfrequency $\omega_n$ into the eigenmode that has a lower eigenfrequency $\omega_k$.

The balance condition (29) can be generalized to an arbitrary case in which several eigenmodes of the atom are excited. Let us make an assumption about the existence of a detailed balance (29) between any two eigenmodes: the emission of energy that is associated with excitation in eigenmodes $n$ and $k$ leads to a redistribution of the electric charge only between those modes. In this case, the derivative $\frac{d|c_n|^2}{dt}$ in equation (29) refers to the rate of change of the amplitude of the eigenmode $n$ in the result of the joint with the eigenmode $k$ emission. Then, the condition of detailed balance (29) means that the decrease in the electric charge in eigenmode $n$ due to emission at a frequency of $\omega_{nk}$ is equal to the increase in the electric charge in eigenmode $k$. The condition of the detailed balance leads to the equation

$$\frac{d|c_n|^2}{dt} = -\sum_{\omega_k<\omega_n}\frac{I_{nk}}{\hbar\omega_{nk}} + \sum_{\omega_k>\omega_n}\frac{I_{nk}}{\hbar\omega_{nk}} \qquad (30)$$

which can also be rewritten in the form

$$\frac{d|c_n|^2}{dt} = -\sum_{\omega_{nk}>0}\frac{I_{nk}}{\hbar\omega_{nk}} + \sum_{\omega_{nk}<0}\frac{I_{nk}}{\hbar\omega_{nk}} \qquad (31)$$

or

$$\frac{d|c_n|^2}{dt} = -\sum_{k\neq n}\frac{I_{nk}}{\hbar\omega_{nk}} \qquad (32)$$

where the summation is taken over all $k \neq n$.

It is easy to show that by summing all of the equations (32) for different values of $n$, one obtains

$$\frac{d}{dt}\sum_n |c_n|^2 = 0 \qquad (33)$$

which is in full agreement with expression (14), while multiplying equation (32) by $\hbar\omega_n$ and summing over all $n$ obtains the energy conservation law (26).

Indeed, let us consider two arbitrary eigenmodes $n$ and $s$ of the electron wave in the hydrogen atom.

For them, equation (32) can be written as

$$\frac{d|c_n|^2}{dt} = -\frac{I_{ns}}{\hbar\omega_{ns}} - \sum_{\substack{k\neq n\\k\neq s}}\frac{I_{nk}}{\hbar\omega_{nk}} \qquad (34)$$

$$\frac{d|c_s|^2}{dt} = \frac{I_{ns}}{\hbar\omega_{ns}} - \sum_{\substack{k\neq s\\k\neq n}}\frac{I_{sk}}{\hbar\omega_{sk}} \qquad (35)$$

Here, we account for the facts $\omega_{sn} = -\omega_{ns}$ and $I_{sn} = I_{ns}$. The sums in the right-hand sides of equations (34) and (35) do not contain the term $(ns)$.

When summing equations (34) and (35), the terms $(ns)$ in the right-hand side will be reduced. Thus, when summing all of the equations in (32), all of the terms on the right-hand side will be reduced; as a result, we obtain expression (33). Similarly, multiplying (34) and (35) by $\hbar\omega_n$ and $\hbar\omega_s$, respectively, and summing obtains

$$\hbar\omega_n \frac{d|c_n|^2}{dt} + \hbar\omega_s \frac{d|c_s|^2}{dt} = -I_{ns} - \omega_n \sum_{\substack{k\neq n \\ k\neq s}} \frac{I_{nk}}{\omega_{nk}} - \omega_s \sum_{\substack{k\neq s \\ k\neq n}} \frac{I_{sk}}{\omega_{sk}} \quad (36)$$

In this case, the sums on the right-hand side do not contain the term $(ns)$. Therefore, if we multiply equation (32) by $\hbar\omega_n$ and calculate the sum, we obtain the energy conservation law (26).

Substituting relation (23) into (32), we obtain

$$\frac{d|c_n|^2}{dt} = -|c_n|^2 \frac{4}{3\hbar c^3} \sum_{k\neq n} \omega_{nk}^3 |c_k|^2 |\mathbf{d}_{nk}|^2 \quad (37)$$

This equation is written for $|c_n|^2$; however, one can assume that it can be extended to the complex amplitudes $c_n$:

$$\frac{dc_n}{dt} = -c_n \frac{2}{3\hbar c^3} \sum_{k\neq n} \omega_{nk}^3 |c_k|^2 |\mathbf{d}_{nk}|^2 \quad (38)$$

Indeed, equation (37) follows from equation (38).

Let us now consider an atom that is in a mixed state with two excited eigenmodes $n$ and $k$, assuming that $\omega_n > \omega_k$. Such an atom will emit electromagnetic waves at a frequency of $\omega_{nk}$. In this case, equation (37) takes the form

$$\frac{d|c_k|^2}{dt} = |c_k|^2 \frac{4}{3\hbar c^3} \omega_{nk}^3 |c_n|^2 |\mathbf{d}_{nk}|^2 \quad (39)$$

$$\frac{d|c_n|^2}{dt} = -|c_n|^2 \frac{4}{3\hbar c^3} \omega_{nk}^3 |c_k|^2 |\mathbf{d}_{nk}|^2 \quad (40)$$

Taking into account that for the case under consideration

$$|c_k|^2 + |c_n|^2 = 1 \quad (41)$$

we obtain

$$\frac{d|c_n|^2}{dt} = -|c_n|^2 (1 - |c_n|^2) \frac{4}{3\hbar c^3} \omega_{nk}^3 |\mathbf{d}_{nk}|^2 \quad (42)$$

and

$$|c_k|^2 = 1 - |c_n|^2 \quad (43)$$

The solution of equation (42) has the form

$$|c_n|^2 = \frac{1}{\frac{1-|c_n(0)|^2}{|c_n(0)|^2} \exp(2\gamma_{nk} t) + 1} \quad (44)$$

and using expression (43), we obtain

$$|c_k|^2 = \frac{1}{\frac{|c_n(0)|^2}{1-|c_n(0)|^2} \exp(-2\gamma_{nk} t) + 1} \quad (45)$$

where

$$\gamma_{nk} = \frac{2}{3\hbar c^3} \omega_{nk}^3 |\mathbf{d}_{nk}|^2 \quad (46)$$

is the damping rate of the spontaneous emission, and $c_n(0)$ is the initial amplitude of the excitation of mode $n$.

Substituting the solutions of (44) and (45) into expression (13), we obtain the wave function of the considered electron wave:

$$\psi = \frac{u_k \exp(-i\omega_k t + i\varphi_k)}{\sqrt{\frac{|c_n(0)|^2}{1-|c_n(0)|^2}\exp(-2\gamma_{nk}t)+1}} + \frac{u_n \exp(-i\omega_n t + i\varphi_n)}{\sqrt{\frac{1-|c_n(0)|^2}{|c_n(0)|^2}\exp(2\gamma_{nk}t)+1}} \qquad (47)$$

where $\varphi_k$ and $\varphi_n$ are the phases of the complex amplitudes $c_k$ and $c_n$, which cannot be found in this approximation. These phases, similar to the modules of the amplitudes, can be slowly varying functions of time.

Because by definition $\gamma_{nk} > 0$, it follows from expressions (44), (45) and (47) that the amplitude of the "upper" mode $n$ decreases with time, and the electron wave will completely go into the "lower" excited eigenmode $k$. This finding means that the degeneration of the mixed state of an atom with two excited modes $n$ and $k$ will occur over time, and the atom will go into a pure state that corresponds to the excited mode $k$.

It follows from expressions (44) and (45) that the characteristic time of the degeneracy of the mixed state $(nk)$ is of the order of $\gamma_{nk}^{-1}$.

Let us estimate the ratio $\gamma_{nk}/\omega_{nk} = \frac{2\omega_{nk}^2}{3\hbar c^3}|\mathbf{d}_{nk}|^2$, while accounting for $|\mathbf{d}_{nk}|\sim ea_B$, $\omega_{nk}\sim\frac{me^4}{\hbar^3}$, where $a_B = \frac{\hbar^2}{me^2}$ is the Bohr radius. As a result, we obtain

$$\gamma_{nk}/\omega_{nk} \sim \alpha^3 \ll 1 \qquad (48)$$

where $\alpha = \frac{e^2}{\hbar c}$ is the fine-structure constant.

As follows from expression (44), $\left|\frac{d|c_n|^2}{dt}\right| < 2\gamma_{nk}$. Then, using (48), we obtain condition (19), which was used in the previous discussion without proof.

From the obtained results, it follows that in the result of the spontaneous emission and the accompanying rearrangement of excited modes, the atom always goes into a pure state that corresponds to the lower of the excited modes. The atom has the lowest eigenmode (ground mode), which corresponds to the lowest eigenfrequency and lowest energy. A pure state that corresponds to this lowest eigenmode is the ground state of the atom. If the atom is in the ground state, then it is possible to excite only the higher eigenmodes. Thus, we arrive at the conclusion that all of the pure states (eigenmodes) of the atom except for the ground state, although they are steady-state, are unstable: even weak excitation of the lower eigenmode leads to "crossflow" of the electron wave (and its electric charge) from an excited mode that has a higher eigenfrequency to an excited mode that has a lower eigenfrequency. At the same time, the ground state of the atom, when only one ground mode is excited, is not only steady-state but also stable: the electron wave always returns to the ground mode from any mixed state in which the ground mode was excited, even weakly.

Note that the solutions in (44) and (45) do not follow from the linear Schrödinger equation, which indicates its limitations. However, they can be obtained from a certain nonlinear equation for the electron wave, which accounts for the inverse action of its emission of an electron wave. The linear approximation of this nonlinear equation is the linear Schrödinger equation. This issue will be considered below.

In the case of weak excitation, which is implemented in most of the experiments under the action of not very intense external influences (e.g., light), we have

$$|c_n|^2 \ll 1, \; 1 - |c_k|^2 \ll 1 \tag{49}$$

In this case, one can use immediately the exact solution in (44) and (45), but one can consider directly equation (42) in this approximation. Let us consider the second way, as it will help to understand (see below) how a probabilistic interpretation of atomic phenomena arises.

Equation (42) in the case of weak excitation (49) of eigenmode $n$ can be approximately written in the form

$$\frac{d|c_n|^2}{dt} = -2\gamma_{nk}|c_n|^2 \tag{50}$$

Note that a weak excitation of mode $n$ can be considered to be a weak perturbation of a pure state that corresponds to eigenmode $k$.

Equation (50) has a solution

$$|c_n|^2 = |c_n(0)|^2 \exp(-2\gamma_{nk}t) \tag{51}$$

and correspondingly, according to the expression in (43), we have

$$|c_k|^2 = 1 - |c_n(0)|^2 \exp(-2\gamma_{nk}t) \tag{52}$$

The wave function of the electron wave in this approximation has the form

$$\psi = u_k \exp(-i\omega_k t) + c_n(0)u_n \exp(-i\omega_n t - \gamma_{nk}t) - \frac{1}{2}|c_n(0)|^2 u_k \exp(-i\omega_k t - 2\gamma_{nk}t) \tag{53}$$

4.4 The width of the spectral lines

According to the results of the previous section, the atom, which was in a mixed excited state, tends monotonically to the pure state, which corresponds to the lower of the excited modes. This continuous "transition" will be accompanied by the attenuating spontaneous emission, and the radiated energy will be equal to the difference between the energy of the atom in the initial (mixed) state and the energy of the atom in the final (pure) state. Although the emission occurs mainly at the carrier frequency $\omega_{nk}$, due to the finite duration of the "transition", other frequencies of the continuous spectrum will be presented in the atom's emission spectrum. As a result, the "spectral line" $\omega_{nk}$ will have a small but finite width.

Let us find the shape of the "spectral line" $\omega_{nk}$ in the approximation of the weak excitation of the upper eigenmode of a mixed state $(nk)$, using expression (53).

Substituting (53) into (15) and considering (19) in the linear approximation with respect to $c_n(0)$, we obtain

$$\ddot{\mathbf{d}} = \mathbf{d}_{nk}[(i\omega_{nk}+\gamma_{nk})^2 c_n(0)\exp(-i\omega_{nk}t - \gamma_{nk}t) + (i\omega_{nk}-\gamma_{nk})^2 c_n^*(0)\exp(i\omega_{nk}t - \gamma_{nk}t)] \quad (54)$$

This equation can be written in the form

$$\ddot{\mathbf{d}} = \frac{1}{2\pi}\int_{-\infty}^{\infty}\ddot{\mathbf{d}}_\omega(\omega)\exp(-i\omega t)\,d\omega \quad (55)$$

where

$$\ddot{\mathbf{d}}_\omega(\omega) = \mathbf{d}_{nk}\left[\frac{(i\omega_{nk}+\gamma_{nk})^2}{-i(\omega-\omega_{nk})+\gamma_{nk}}c_n(0) + \frac{(i\omega_{nk}-\gamma_{nk})^2}{-i(\omega+\omega_{nk})+\gamma_{nk}}c_n^*(0)\right] \quad (56)$$

Accounting for the assessment in (48) as well as the condition $\omega_{nk} > 0$, it is clear that the second term in the brackets will be of the order of $\alpha^3$ compared to the first term. In the approximation under consideration, it can be neglected. Then, one can write approximately

$$\ddot{\mathbf{d}}_\omega(\omega) = c_n(0)\mathbf{d}_{nk}\frac{(i\omega_{nk}+\gamma_{nk})^2}{-i(\omega-\omega_{nk})+\gamma_{nk}} \quad (57)$$

The electromagnetic energy emitted by the electron wave during all of the time of the spontaneous emission of the excited hydrogen atom is

$$\mathcal{E} = \int_0^\infty I\,dt \quad (58)$$

Substituting (21) and considering (55), we obtain [23]

$$\mathcal{E} = \frac{4}{3c^3}\int_0^\infty |\ddot{\mathbf{d}}_\omega(\omega)|^2\frac{d\omega}{2\pi} \quad (59)$$

As a result, we can find the distribution on the spectrum of the energy emitted in the spontaneous emission of an atom that is in a mixed excited state $(nk)$ with a weakly excited upper eigenmode $n$. According to (59), the energy that is emitted in the frequency range $[\omega, \omega + d\omega]$ is equal to

$$d\mathcal{E}_\omega = \frac{4}{3c^3}|\ddot{\mathbf{d}}_\omega(\omega)|^2\frac{d\omega}{2\pi} \quad (60)$$

Accounting for (57), we obtain

$$d\mathcal{E}_\omega = \frac{4\omega_{nk}^4}{3c^3}|\mathbf{d}_{nk}|^2\frac{|c_n(0)|^2}{(\omega-\omega_{nk})^2+\gamma_{nk}^2}\frac{d\omega}{2\pi} \quad (61)$$

This expression determines the shape of the spectral line $\omega_{nk}$. The result was obtained in the framework of classical electrodynamics using only classical representations and coincides with the shape of the spectral line, which is predicted by quantum electrodynamics using the method of second quantization [24].

## 5 Nonlinear Schrödinger equation that describes the spontaneous emission

The linear Schrödinger equation allows us to describe the source of a spontaneous emission, which is completely described by classical electrodynamics, but it cannot describe the change in the electron wave that occurs during the spontaneous emission, due to the loss of energy. To accomplish this goal, we had to use additional considerations based on the law of conservation of energy, which allowed us to describe both the changes that occur with the electron wave itself in the process of spontaneous emission and the spontaneous emission characteristics that are usually described within the framework of quantum electrodynamics.

Obviously, such a situation is not satisfactory. It appears that the above results should follow from the theory itself without the involvement of any additional, albeit plausible, considerations. Let us consider equation (50) while accounting for (46) for the case of weak excitation of eigenmode $n$, when condition (49) is satisfied:

$$\frac{d|c_n|^2}{dt} = -|c_n|^2 \frac{4}{3\hbar c^3} \omega_{nk}^3 |\mathbf{d}_{nk}|^2 \tag{62}$$

This equation can be obtained from the equation

$$\frac{dc_n}{dt} = -c_n \frac{2}{3\hbar c^3} \omega_{nk}^3 |\mathbf{d}_{nk}|^2 \tag{63}$$

In this approximation

$$\frac{dc_k}{dt} \approx 0 \tag{64}$$

Using definition (18), equation (63) can be written as

$$\frac{dc_n}{dt} = c_n \frac{2e\omega_{nk}^3}{3\hbar c^3} \mathbf{d}_{nk} \int \mathbf{r} u_n^*(\mathbf{r}) u_k(\mathbf{r}) dV \tag{65}$$

Considering (12), this equation is equivalent to

$$\frac{dc_n}{dt} u_n = c_n \frac{2e\omega_{nk}^3}{3\hbar c^3} u_k \mathbf{r} \mathbf{d}_{nk} \tag{66}$$

Let us consider the wave function of an electron wave that has two excited eigenmodes $k$ and $n$:

$$\psi = c_k u_k \exp(-i\omega_k t) + c_n u_n \exp(-i\omega_n t) \tag{67}$$

Let us differentiate (67) with respect to time, while considering equations (64) and (66); at the same time, we account for the fact that the frequencies $\omega_k$ and $\omega_n$ are eigenfrequencies, and the functions $u_k$ and $u_n$ are the corresponding eigenfunctions of the linear Schrödinger equation for the hydrogen atom. As a result, we obtain the equation

$$i\hbar \frac{\partial \psi}{\partial t} = -\frac{1}{2m}\Delta\psi - e\varphi_p\psi + i\frac{2e\omega_{nk}^3}{3c^3}|c_k|^2 u_k c_n \mathbf{r}\mathbf{d}_{nk} \exp(-i\omega_n t) \tag{68}$$

where $\varphi_p$ is the scalar potential of the electrostatic field of the proton.

It is easy to check that equations (63) and (64) definitely follow from this equation in approximation (49).

Using (11) and (18), we can write equation (68) in the form

$$ih\frac{\partial\psi}{\partial t} = -\frac{1}{2m}\Delta\psi - e\varphi_p\psi - i\frac{2e^2\omega_{nk}^3}{3c^3}(c_k\psi_k)\mathbf{r}\int(c_k\psi_k)^*\mathbf{r}(c_n\psi_n)dV \tag{69}$$

while considering that $\psi_k^*\psi_n$ depends on time as in $\exp(-i\omega_{nk}t)$ and the fact that in accordance with (48) $\gamma_{nk} \ll \omega_{nk}$, equation (69) can be rewritten as

$$ih\frac{\partial\psi}{\partial t} = -\frac{1}{2m}\Delta\psi - e\varphi_p\psi - \frac{2e^2}{3c^3}(c_k\psi_k)\mathbf{r}\frac{\partial^3}{\partial t^3}\int(c_k\psi_k)^*\mathbf{r}(c_n\psi_n)dV \tag{70}$$

With the same accuracy (up to the first order for a small $c_n$), considering (19), we can write

$$ih\frac{\partial\psi}{\partial t} = -\frac{1}{2m}\Delta\psi - e\varphi_p\psi - \frac{2e^2}{3c^3}\psi\mathbf{r}\frac{\partial^3}{\partial t^3}\int(c_k\psi_k)^*\mathbf{r}\psi d\mathbf{r} \tag{71}$$

Let us consider the equation

$$ih\frac{\partial\psi}{\partial t} = -\frac{1}{2m}\Delta\psi - e\varphi_p\psi - \frac{2e^2}{3c^3}\psi\mathbf{r}\frac{\partial^3}{\partial t^3}\int \mathbf{r}|\psi|^2 d\mathbf{r} \tag{72}$$

and show that equation (72) in this approximation is equivalent to equation (71). Substituting (67) into equation (72) and neglecting the terms of the order of $\gamma_{nk}/\omega_{nk} \ll 1$ and higher, we obtain up to the small of the order of $c_n$

$$ih\frac{\partial\psi}{\partial t} = -\frac{1}{2m}\Delta\psi - e\varphi_p\psi + i\frac{2e\omega_{nk}^3}{3c^3}|c_k|^2 u_k c_n\mathbf{r}\mathbf{d}_{nk}\exp(-i\omega_n t)[1 - (c_k/c_k^*)(c_n^*/c_n)\exp(i2\omega_{nk}t)] \tag{73}$$

This equation is different from the original equation (68) by a factor of the latter term, which oscillates around the unit with a frequency $2\omega_{nk}$. This oscillating term will play the role of an external periodic perturbation. Such an action will be weak, and it can be analyzed within perturbation theory. According to perturbation theory [17], the only eigenmodes that will be excited noticeably are those whose frequency coincides with or is very close to the frequency of the external influence. For this reason, the second term in brackets in equation (73) can be discarded, and equation (73) will coincide with the original equation (68). This arrangement means that equations (63) and (64), which describe spontaneous emission at a weak excitation of eigenmode $n$, follow from equation (72).

Accounting for (15), equation (72) can be written as

$$ih\frac{\partial\psi}{\partial t} = -\frac{1}{2m}\Delta\psi - e\varphi_p\psi + \frac{2e}{3c^3}\mathbf{r}\ddot{\mathbf{d}}\,\psi \tag{74}$$

The last term in the right-hand side of equation (74) has a simple interpretation. According to classical electrodynamics, the electric-dipole-radiating system creates around itself an additional (induced) electromagnetic field with a vector potential [23]

$$\mathbf{A}_r = -\frac{2}{3c^2}\ddot{\mathbf{d}} \tag{75}$$

while the corresponding scalar potential is equal to zero:

$$\varphi_r = 0 \tag{76}$$

Because $\mathbf{A}_r$ does not depend on the coordinates, an additional magnetic field is zero

$$\mathbf{H}_r = \text{rot}\, \mathbf{A}_r = 0 \tag{77}$$

while the additional electric field will be determined by the usual relationships of classical electrodynamics [23], which in this case takes the form

$$\mathbf{E}_r = -\frac{1}{c}\frac{\partial \mathbf{A}_r}{\partial t} \tag{78}$$

or using (75)

$$\mathbf{E}_r = \frac{2}{3c^3}\dddot{\mathbf{d}} \tag{79}$$

The additional electric field (79) that is generated by the electric-dipole-radiating system has an inverse action on the radiating system. In classical electrodynamics with point-charged particles, this arrangement leads to the appearance of an additional Lorentz-Abraham force [23]

$$\mathbf{f} = -\frac{2e}{3c^3}\dddot{\mathbf{d}} \tag{80}$$

Thus, the electric-dipole-radiating electron wave in the hydrogen atom is in the electrostatic field of the nucleus and in the additional self-radiation field, which is described by the vector and scalar potentials (75) and (76). In this case, the Schrödinger equation takes the form

$$i\hbar \frac{\partial \psi}{\partial t} = \left[\frac{1}{2m}\left(\frac{\hbar}{i}\nabla + \frac{e}{c}\mathbf{A}_r\right)^2 - e\varphi_p\right]\psi \tag{81}$$

This equation can be rewritten in another, more convenient form for analysis, while accounting for the gauge invariance of the Maxwell-Schrödinger field: the joint system of the Maxwell-Schrödinger equations does not change its form under the gauge transformations [17,23]

$$\mathbf{A} \to \mathbf{A} + \nabla f, \;\; \varphi \to \varphi - \frac{1}{c}\frac{\partial f}{\partial t}, \;\; \psi \to \psi \exp(-\frac{ie}{\hbar c}f) \tag{82}$$

where $f$ is an arbitrary scalar function.

In particular, by choosing

$$f = \frac{2}{3c^2}\mathbf{r}\ddot{\mathbf{d}} \tag{83}$$

we obtain, for the potentials of the total electromagnetic field acting on the electron wave in the hydrogen atom, the following:

$$\mathbf{A}' = 0, \;\; \varphi' = \varphi_p + \varphi_r \tag{84}$$

where

$$\varphi_r = -\frac{2}{3c^3}\mathbf{r}\dddot{\mathbf{d}} \tag{85}$$

Substituting (84) and (85) into the Schrödinger equation, we obtain

$$i\hbar \frac{\partial \psi}{\partial t} = -\frac{1}{2m}\Delta\psi - e\varphi'\psi \tag{86}$$

which coincides with equation (74) obtained above. It follows that the last term in the Schrödinger equation (72), (74) accounts for the inverse action of the spontaneous emission of the electron wave.

Note that although we have equations (74) and (81), (75) describes the same electron wave in the hydrogen atom, and their solutions do not coincide; they are related by the expression

$$\psi' = \psi \exp(-\frac{i2e}{3\hbar c^3}\mathbf{r}\ddot{\mathbf{d}}) \qquad (87)$$

where $\psi$ is the solution of equation (81), and (75) while $\psi'$ is the solution of equation (74). Solutions $\psi$ and $\psi'$ are different by the rapidly oscillating phase $(-\frac{2e}{3\hbar c^3}\mathbf{r}\ddot{\mathbf{d}})$.

Thus, the spontaneous emission of the hydrogen atom is completely described by the Maxwell-Schrödinger system of equations, in which the electromagnetic and electron fields are considered to be classical and self-radiation fields, created by the electron wave that is accounted for. The presence of the last term in equations (72), (74) makes it nonlinear. It is this nonlinearity of equations (72) and (74) allows describing the spontaneous emission of an atom, and (as it will be shown in the subsequent papers of this series) other "quantum" phenomena that cannot be described by the linear Schrödinger equation and that are traditionally described within quantum electrodynamics.

## 6  Statistical interpretation of spontaneous emission

As was shown above, for an explanation of the basic atomic regularities, there is no need for the quantization of energy (neither of an atom nor of an electromagnetic field). We have seen that the particles ("photons" and "electrons"), as physical objects, are completely superfluous for explaining the processes that occur in the atom. For their description, it is sufficient to have classical electrodynamics and the understanding that instead of an "electron", we should consider an electrically charged electron wave that represents an unusual but classical field. The analysis above has shown that all of the processes (at least those discussed above) are completely deterministic and do not require the involvement of the apparatus of probability theory.

With respect to this connection, it is interesting to understand why a statistical description of atomic phenomena that are accepted in quantum mechanics and quantum electrodynamics allows obtaining the correct results.

Observing discrete events (the clicks of a detector, spots on a photographic plate, the discrete spectrum of the spontaneous emission of an atom) that are induced by some continuous process (e.g., a wave), the temptation arises to explain them by a discreteness in the process itself, i.e., to represent a continuous process as a sequence of discrete events, or quanta. Examples are the experiments with so-called "single photons" or "single electrons" (primerely in the double-slit experiments), which have an absolutely classical, non-quantum explanation [13-16], although precisely the wave-particle ("quantum") interpretation is now common and popular. Applying the "quantum" interpretation to a non-quantum (continuous) process, we will certainly face internal contradictions for eliminating which we will be forced to turn into a probabilistic description. In other words, we will have to replace the actual continuous deterministic process by a fictitious discrete random process and to determine the probabilities of the discrete events to preserve the observed effects.

This approach fully applies to the quantum mechanical explanation of the discrete spectra of the spontaneous emission of atoms, which dates back to the mechanistic Bohr theory.

Let us consider this approach in more detail.

As was shown above, an atom being in a mixed state (13) emits electromagnetic waves of the discrete spectrum $\omega_{nk} = \omega_n - \omega_k$ in full compliance with classical electrodynamics. Thus, *all of the frequencies $\omega_{nk}$ that correspond to the excited eigenmodes of the atom will be presented simultaneously in the emission spectrum of each atom while it is in a mixed state (13)*. This arrangement can be seen if we observe the emission from a single atom in which several eigenmodes are excited simultaneously (this state can be achieved, e.g., by irradiating the atom with an electromagnetic wave that contains several characteristic frequencies that correspond to the eigenfrequencies of the atom, $\omega_n$). Such a conclusion is fundamentally different from the traditional interpretation that is accepted in quantum mechanics, according to which every atom that is in an excited state $n$ can simultaneously make only one discrete quantum transition $n \to k$ to a lower energy level $k$, during which a photon that corresponds to one spectral line $\omega_{nk}$ will be emitted; at the same time, the entire emission spectrum of an atom can be observed in experiments because we are addressing a large number of atoms that are in different "quantum states", which make the different quantum transitions.

According to expression (23), the intensity of each spectral line $\omega_{nk}$ depends simultaneously on the amplitudes of excitation of both eigenmodes $|c_n|^2$ and $|c_k|^2$.

Let us consider the case in which several eigenmodes are simultaneously excited in the atom, and we assume that all of the modes, except for the lower of the excited states, which we denote by the index $k$ (this mode is not necessarily the ground mode), are excited weakly. This circumstance means that for all $n \neq k$, condition (49) is satisfied. In this case, according to (23), the intensity of the atomic emission at the frequency $\omega_{nk}$ is approximately determined by the expression

$$I_{nk} = \frac{4\omega_{nk}^4}{3c^3}|c_n|^2|\mathbf{d}_{nk}|^2 \tag{88}$$

The intensity of the emission at the frequency $\omega_{n_1 n_2}$ corresponds to any two weakly excited modes $n_1$ and $n_2$ that are small of the order of $|c_{n_1}|^2|c_{n_2}|^2 \ll 1$, and they will be neglected. In this case, only the lines $\omega_{nk}$ that correspond to different $n$ will be clearly observed in the emission spectrum of an atom.

Let us consider an ensemble that consists of $N$ identical atoms that are in exactly the same mixed states; we assume that the amplitudes of the excitation of the same eigenmodes $|c_n|^2$ of all of the atoms of the ensemble are the same. Then, the total intensity of the emission of such an ensemble of atoms at the frequency $\omega_{nk}$ will be

$$(I_{nk})_N = N|c_n|^2 \frac{4\omega_{nk}^4}{3c^3} |\mathbf{d}_{nk}|^2 \tag{89}$$

Because the value $-e|c_n|^2$ is equal to the charge of the electron wave, which is "concentrated" in eigenmode $n$ of the atom, the total electric charge of the electron waves, "concentrated" in mode $n$ of all of the atoms of the ensemble, will be equal to

$$(q_n)_N = -Ne|c_n|^2 \tag{90}$$

This expression was obtained within classical electrodynamics without the use of any particles; however, we can give it a statistical interpretation in the spirit of quantum mechanics. Let us assume that each atom can be in only one excited state $n$, where the "electron", which is considered to be a particle that has the charge $-e$, is located on the "orbital" $n$. Let $N_n$ be the number of atoms in the excited state $n$ in the considered ensemble. Then, the total charge of the electrons of all of these atoms will be $(q_n)_N = -eN_n$. To have this interpretation give the same result as classical electrodynamics, this value $(q_n)_N$ should coincide with (90). As a result, we obtain $N_n = N|c_n|^2$. Thus, in the corpuscular interpretation, the value $|c_n|^2$ should be considered to be a fraction of the atoms of the ensemble that is in the discrete state $n$. If we randomly choose an atom from this ensemble, then this atom will be in state $n$ with a probability $p_n = |c_n|^2$. Thus, we arrive at the probabilistic interpretation of the parameter $|c_n|^2$, according to which $|c_n|^2$ is the probability that the atom is in the excited state $n$. In this case, expression (14) should be considered to be a normalization condition for the probabilities in full accordance with the interpretation that is accepted in quantum mechanics.

Let us assume that an atom that is in an excited state $n$ makes a spontaneous "quantum jump" to the lower energy level $k$ and that this jump is accompanied with an emission of an energy quantum $\hbar\omega_{nk}$. To describe such transitions, we must enter the probabilities. Let the probability of the quantum spontaneous transition $n \to k$ per unit time be equal to $w_{nk}$. Then, the energy that is emitted by an ensemble of atoms at a frequency of $\omega_{nk}$ per unit time (i.e., the total intensity of the emission), will be

$$(I_{nk})_N = N\hbar\omega_{nk} p_n w_{nk} \tag{91}$$

The corpuscular interpretation of the spontaneous emission will give the same result as that from classical electrodynamics, if expressions (89) and (91) coincide. As a result, we obtain

$$w_{nk} = \frac{4\omega_{nk}^3}{3\hbar c^3}|\mathbf{d}_{nk}|^2 \tag{92}$$

which coincides with a well-known result from quantum electrodynamics [24].

Let us now consider the "kinetics" of "quantum transitions" in the corpuscular interpretation. In this case, one can write the kinetic equation for the number of atoms that are in the excited state $n$, which is the usual equation for random decay:

$$\frac{dN_n}{dt} = -w_{nk}N_n \tag{93}$$

Accounting for the definition of $N_n$ and the expressions (92) and (46), we obtain equation (50), which was actually obtained in the framework of classical field theory without any quantization. Thus, we see that it is indeed possible to give a statistical corpuscular interpretation to the results obtained within the classical field theory if we consider the approximation of a weak excitation of the upper eigenmodes of the atom, although the artificiality of such an interpretation is now obvious.

# 7   Concluding remarks

Enumerate the main results of this paper.

1. It is shown that the hydrogen atom can be considered to be a classical open volume resonator in which the electron wave - a classical wave field - is held by the electrostatic field of the nucleus.

2. The hydrogen atom and its basic properties, primarily the regularities of spontaneous emission, are fully described within classical field theory without any quantization of the electromagnetic and electron fields. In particular, it is shown that there are no jump-like transitions in the atom, and the spontaneous emission occurs not in the form of discrete portions - quanta - but continuously during all of the time as long as the atom is in the mixed excited state. The spontaneous emission of a single atom occurs simultaneously on all frequencies and corresponds to all pairs of excited eigenmodes. When at least two eigenmodes are excited in an atom, the spontaneous emission in full compliance with classical electrodynamics begins, without any delay and without an external "push". For this reason, to answer the question of why a spontaneous emission occurs, there is no need to create a hypothesis such as a "trembling of the electron caused by fluctuations of QED vacuum" (so-called Zitterbewegung). Moreover, there is no need to introduce a concept such as a QED vacuum. There is also no need to create so-called stochastic electrodynamics because all of the above considered processes in the atom are completely deterministic and Maxwell electrodynamics is sufficient for their description.

3. An explanation of the corpuscular-statistical interpretation of the atomic phenomena has been given. It was shown that this interpretation is only a misinterpretation of a deterministic continuous process, which is based on the formal similarity of the equations that describe the atom at a weak excitation of the upper eigenmodes and the equations that describe a Markov process of transition of an abstract system between the discrete states. Furthermore, at a strong excitation of the upper eigenmodes, when condition (49) is not satisfied, the corpuscular-statistical theory will give incorrect results.

4. In the framework of classical electrodynamics, the nonlinear Schrödinger equation was obtained, which accounts for the inverse action of self-electromagnetic radiation of the electron wave; this equation is sufficient for the description of all of the regularities of the spontaneous emission of the hydrogen atom.

Obviously, many other properties of the hydrogen atom, which are considered in conventional courses on quantum mechanics (e.g., Stark effect, normal Zeeman effect), can be completely transferred into the proposed theory without any change (because they were obtained by the formal solution of the Schrödinger equation), but already without the corpuscular interpretation.

Note that although this theory has some similarities to semiclassical theories [8-12], it is by no means a semiclassical. The present theory is a *fully classical theory*! Indeed, in semiclassical theories, light is considered to be a classical electromagnetic wave, but the electrons are considered to be quantum particles that are described by wave equations, for example, the Schrödinger equation. In semiclassical theories, an atom as before is interpreted as a system that has discrete energy levels and is able to suddenly transit from one stationary level to another. In contrast, there are not any particles (neither photons nor electrons) in the proposed theory, but instead, there are only classical continuous waves – the electromagnetic and electron, and the energy of the atom changes continuously.

The obtained results show that quantum mechanics should be considered not as a development of mechanics but as a pure field theory in the spirit of Maxwell electrodynamics. In other words, quantum mechanics is not a theory of particles but is instead a classical theory of wave fields, whose behaviour is similar to the behaviour of the Maxwell field. For this reason, the interpretation of quantum mechanics should be made not in terms of mechanics: particles, jump-like transitions, momentum, and energy, but in terms of field theory: frequency, wavelength, distributions in the space parameters (e.g., energy, momentum, electric charge).

The results of this paper and the papers in [13-16] show that considering quantum theory to be a theory of classical fields, we will not face difficulties in the interpretation of its results, including when comparing the results of the theory with experimental data.

However, this paper proposes a new problem: why does the electron wave not feel its own electrostatic field? Indeed, according to section 5, the nonlinear Schrödinger equation for the electron wave in a hydrogen atom must have the form

$$i\hbar \frac{\partial \psi}{\partial t} = \left[ \frac{1}{2m} \left( \frac{\hbar}{i} \nabla + \frac{e}{c} \mathbf{A} \right)^2 - e(\varphi - \varphi_0) \right] \psi \tag{94}$$

where the potentials $(\varphi, \mathbf{A})$ describe the total electromagnetic field (both the field of the nucleus and the self-electromagnetic field of the electron wave), while $\varphi_0$ satisfies the equation

$$\Delta \varphi_0 = 4\pi e |\psi|^2 \tag{95}$$

even for the unsteady field $\psi$. The function $\varphi_0$ can be considered to be the potential of the "electrostatic" field that is created by the charged electron wave in the hydrogen atom. At the same time, the potentials $(\varphi, \mathbf{A})$ of the total electromagnetic field are described by the well-known expressions of classical electrodynamics [23]

$$\varphi = \varphi_p - e \int \frac{|\psi_{t-R/c}|^2}{R} dV \tag{96}$$

$$\mathbf{A} = \frac{1}{c} \int \frac{\mathbf{j}_{t-R/c}}{R} dV \tag{97}$$

where **j** is the electric current density of the electron wave, which is defined by the well-known expression of quantum mechanics [16,17].

It is easy to show [23] that for terms up to the order of $1/c^3$ in the expansion of the scalar potential (96), equations (94)-(97) turn into equation (86) and, consequently, to equations (72) (74), which account for the spontaneous emission of an atom.

From equation (94), it follows that the electrostatic field of the nucleus and the electrostatic field of the electron wave are unequal in their actions on the electron wave: the electrostatic field of the nucleus acts on it, while its own electrostatic field does not act on it.

It remains unexplained why the electron field, which has the electric charge and, therefore, creates an electromagnetic field, does not "feel" its own electrostatic field, although it "feels" its own unsteady (radiation) field.

This issue is expected to be considered in the subsequent papers of this series. In my opinion, after we answer this question, it will be easy to construct a classical field-theory of "multi-electron" atoms, which would be similar to that described above for the hydrogen atom. The answer to this question will also allow us to write a relativistic wave equation (the Dirac equation) for the electron wave while accounting for the inverse action of its own radiation.

In the subsequent papers of this series, it will be shown that also other "quantum" phenomena, including phenomena associated with spin, can be explained without any quantization in the framework of classical field theory.

**Acknowledgments**

Funding was provided by Tomsk State University competitiveness improvement program.